# Towards a Security Baseline for IaaS-Cloud Back-Ends in Industry 4.0

Elisabeth Bauer, Oliver Schluga, Silia Maksuti, Ani Bicaku, David Hofbauer, Igor Ivkic, Markus G. Tauber
University of Applied Sciences Burgenland
Eisenstadt, Austria

Alexander Wöhrer
University of Vienna
Vienna, Austria

*Abstract*—The popularity of cloud based Infrastructure-as-a-Service (IaaS) solutions is becoming increasingly popular. However, since IaaS providers and customers interact in a flexible and scalable environment, security remains a serious concern. To handle such security issues, defining a set of security parameters in the service level agreements (SLA) between both, IaaS provider and customer, is of utmost importance. In this paper, the European Network and Information Security Agency (ENISA) guidelines are evaluated to extract a set of security parameters for IaaS. Furthermore, the level of applicability and implementation of this set is used to assess popular industrial and open-source IaaS cloud platforms, respectively VMware and OpenStack. Both platforms provide private clouds, used as backend infrastructures in Industry 4.0 application scenarios. The results serve as initial work to identify a security baseline and research needs for creating secure cloud environments for Industry 4.0.

## I.   INTRODUCTION

Today's working environment is affected by continuous and rapid change, cost saving aspects and the need for flexibility. This influences social as well as technical environments. Cloud computing offers possibilities to handle the technical challenges while being steadily developed. It is defined by the National Institute of Standards and Technology (NIST) as a model for enabling ubiquitous, convenient, on-demand network access to a shared pool of configurable computing resources (e.g., networks, servers, storage, applications, and services) that can be rapidly provisioned and released with minimal management effort or service provider interaction. The three cloud service delivery models according to NIST are: Software as a Service (SaaS), Platform as a Service (PaaS) and Infrastructure as a Service (IaaS) [1].

The focus of this paper lies on IaaS, which represents the fundamental building blocks for cloud services. IaaS is composed of highly automated and scalable compute resources, complemented by cloud storage and network capability, which can be self-provisioned, metered, and available on-demand [1]. This is also used to provide private clouds, which can have impact on industry, where cloud computing is a part of the proposed concept for the 4th industrial revolution, referred to as Industry 4.0. In particular in this domain stakeholder built on dedicated infrastructure rather than on public clouds. Dedicated infrastructure can be based within a trusted providers premises or independent divisions in large enterprises.

The cloud servers and their corresponding resources offered by IaaS providers can be accessed via dashboards or APIs. Furthermore, IaaS customers have direct access to their servers and storage. This implies higher order of scalability and lower investment in capacity planning, physical maintenance and management. Compared to other cloud computing models, IaaS is the most flexible, which allows automated deployment of servers, processing power, storage, and networking. IaaS customers have true control over their infrastructure, compared to customers of PaaS or SaaS services. Thus, IaaS is the most used cloud service model, respectively by cloud service providers and customers [2].

As stated above both, IaaS providers and users, interact in a flexible and scalable environment. To handle organizational and legal aspects, SLA are set up to define clear Service Level Objectives (SLO), to measure the service performance and to deal with security issues, which can be very difficult to define. Therefore, measurable security related SLOs are needed to give a clear definition of the covered security aspects. To address this issue towards a security base line for IaaS platforms in an Industry 4.0 environment, the ENISA guidelines are used to extract measurable security related SLO metrics. The measurable security related SLOs act as an input to assess popular industrial cloud platforms, which provide a lot of features, and to assess popular open source cloud platform, which provide a number of customized services. In this paper, the platforms considered for assessment are the VMware vCloud Suite and OpenStack. Moreover, both are compared to highlight the main differences concerning security related SLOs.

Thus, the contribution of this paper is twofold:

- first, we identify a set of topics that still require research and development and

- secondly, we provide a comparison of popular industrial and open-source platforms focusing on private cloud environments, which are important for Industry 4.0 use cases.

## II. RELATED WORK

Related research work has shown the correlation between security standards and SLAs for cloud infrastructure. For instance, the benefits and costs of introducing SLA matching mechanisms for cloud infrastructures are discussed in [3], but there is no explicit focus on security and IaaS in Industry 4.0 environments. Besides that, in [4] the main security concerns and solutions are identified, while [5] presents security and privacy requirements referring to ISO, NIST and CSA. To show the applicability of these three standards, Hoehl [6] compares their service models, whereas the International Telecommunication Union (ITU) analyzed the current standards and guidelines concerning cloud security in [7].

Within these papers, security related SLAs are defined and discussed generically. The focus lies on the contractual relationship between the cloud service provider and the customer and the specified responsibilities between them. Another outcome of these papers is defining common metrics for measuring security service levels. Rathbun [8] shows the importance of security metrics and answers general questions addressing the reasons for having security metrics. Based on this outcome, it is possible to generate the essential information concerning cloud security assessment for our research work.

The CSA has published a new version of the Cloud Controls Matrix. An overview of all common standards and legal publications concerning cloud computing, including ENISA, ISO, NIST, BSI and more are mentioned in [9]. These metrics give an overview of the current state of the art concerning the offer of security standards and guidelines within cloud computing, but no platform's assessment.

Other related work assesses IaaS solutions such as [2], where security vulnerabilities from inside and outside concerning OpenStack are discussed, but without the consideration of ENISA guidelines.

The related work mentioned above consists of general information about security metrics, guidelines and standards as well as comparative aspects between them. Furthermore, it includes the assessment of different cloud computing offers. This paper relies on ENISA guidelines for monitoring of security in cloud contracts (PRO-SEC) [10]. Based on this guideline the extraction of ENISA specific needs regarding SLOs from the cloud service provider's point of view are discussed. This paper focuses on the analysis of the applicability and implementation of the extracted security SLOs on the industrial software suite VMware vCloud and the open source software services of OpenStack. The comparison of both analyses shows the main differences concerning security of both solutions.

To the best of our knowledge, no publication was identified that evaluates IaaS solutions based on ENISA criteria, comparing VMware to OpenStack cloud platforms.

## III. SECURITY RELATED BASELINES

In the area of security, a variety of standards and guidelines concerning cloud computing are in place. This section, gives a brief overview of these standards and guidelines.

### A. International Organization for Standardization

The ISO standards cover a wide spectrum of topics which are generated out of market needs. Security is covered by the ISO 2700x family, which provides an overview and explains the Information Security Management Systems (ISMS), referring to ISMS family of standards with related terms and definitions. The most relevant standards of ISO 2700x family addressing cloud computing security are listed below:

- ISO/IEC 27001 focuses on the requirements for Information Security Management Systems (ISMS) [11]
- ISO/IEC 27002 provides guidelines and general principles for organizations and their ISMS [12]
- ISO 27017 generally focuses on the protection of the in-formation in the cloud services and is built upon the existing security controls of ISO 27002.

The ISO 2700x family provides a set of security criteria and addresses also cloud technologies such as, ISO 27017. Therefor it provides a comparable basis for the further extraction of security criteria within the ENISA guidelines.

### B. National Institute of Standards and Technology (NIST)

NIST is responsible for developing information security standards and guidelines, including minimum requirements for federal information systems. The most relevant guidelines of NIST addressing cloud computing security are listed above:

- NIST Special Publication 800-53 provides as set of security controls which fulfill the security requirements based on organizations, business processes and information systems, including a risk management framework. Additionally, it provides a set of security controls [13].
- NIST Special Publication 800-144 provides an overview of public cloud computing and guidelines for protecting security and privacy [14].

The NIST publications provide comparable security controls and the setup of the NIST Cloud Computing Program (NCCP) stands for further development in this area. Nevertheless, within this research work the status quo in the year 2016 is taken into account.

### C. Cloud Security Alliance

The CSA security guidance working group provides 14 do-mains about critical areas of cloud computing and works on continuous improvement of the published best practice methods concerning security. The CSA itself is member driven and provides best practice efforts.

### D. Bundesamt für Sicherheit in der Infromationstechnik

The BSI Germany provides catalogues concerning IT Security Management and golden rules for application. These catalogues mainly have an organizational view on security and no direct recommendation for the evaluation of IaaS offers. The BSI provides minimum information security requirements for Cloud Service provider.

## E. European Commission Guideline

The European Commission (EU) has published in 2014 the "Cloud Service Level Agreement Standardization Guideline" (C-SIG). This guideline focuses on the international orientated standardization of the contractual relationship between a cloud service provider and a cloud service customer. EU aims to make it easier for cloud services customers to compare cloud services offered by different providers by using standardized SLAs. Using a standard set of metrics in a cloud SLA makes it easier and faster to define the SLA. This also simplifies the task comparing one cloud SLA to another. The SLOs covered in C-SIG are: (i) Performance SLOs, (ii) Security SLOs, (iii) Data Management SLOs, and (iv) Personal Data Protection SLOs. The definition of SLOs in C-SIG relies on standards and guide-lines produced by organizations such as ENISA, NIST or ISO/IEC.

## F. European Network and Information Securtiy Agency

ENISA is defined as a center of expertise for cyber security in Europe, build up in 2004. The agency's role is to guide experts towards security solutions that are adapted to the needs of the internal market. Therefore, the development of security approaches and solutions that are interoperable across the whole EU is one of the agency's aims. In general, guidelines and recommendations published by ENISA are quite unknown within the ITIL community.

In contrast to ISO, NIST, CSA and BSI, ENISA focuses on the European market and is not a member driven organization such as CSA. In the comparison of the standards, none of them includes unique criteria. Moreover, the standards have many similarities in the provided criteria. Due to that fact that ENISA criteria refers to guidelines and standards mentioned above, the criteria is not very different. Procure secure (PRO-SEC) [10] is a guideline, which concentrates on the reporting and alerting of key measurable parameters, as well as a clear understanding of how to manage the customer's own responsibilities for security. This customer centric approach provides the basis for understanding security risks, which in combination with measurable parameters shows the special focus of ENISA. Due to these aspects, ENISA is used to extract a set of security related SLOs described in the following section.

## IV. ENISA EVALUATION

As mentioned before, ENISA has published a guide for monitoring security including service availability and continuity in SLAs (PRO-SEC) [10]. This security monitoring framework defines the following parameter groups (P.G.):

- Service Availability (SA)
- Incident Response (IR)
- Service Elasticity and Load Tolerance (SELT)
- Data Life Cycle Management (DLCM)
- Technical Compliance and Vulnerability Management (TCVM)
- Change Management (CM)
- Data Isolation (DI)
- Log Management and Forensics (LMF)

Table I lists the ENISA P.G.s, together with a short description, a possible monitoring solution and a set of software-defined-datacenter (SDDC) applicable security related SLOs which have been extracted from C-Sig [15]. A SDDC is a data storage facility where all elements of the infrastructure, networking, storage, CPU and security are virtualized and delivered as a service. The entire process from deployment to monitoring and automation of the infrastructure is abstracted from hardware and implemented in software. Typically, the core architectural components that comprise a SDDC include the following [16]: (i) Computer virtualization (compute), (ii) Software-defined storage (SDS), (iii) Software-defined network (SDN), and (iv) Management and automation software (business logic layer - BLC).

In the next section, the applicability and implementation of this set of security related SLOs within VMware's SDDC and OpenStack is discussed.

TABLE I. ENISA P.G. AND SECURITY RELATED SLOS

| P.G. | Description (D) & Monitoring (M) | Security related SLOs |
|---|---|---|
| SA | **D:** clearly defines when a service is considered available and which functions and services are covered by availability monitoring. **M:** log examination running service health-checks using monitoring tools | **SA01** Service uptime **SA02** Percentage of successful requests **SA03** Percentage of timely service provisioning requests **SA04** Average response time **SA05** Maximum response time **SA06** Level of redundancy **SA07** Service reliability |
| IR | **D:** relates to how a cloud service provider responds to and recovers from incidents. **M:** collecting log data about time related events, e.g. incident reporting | **IR01** Support hours **IR02** Support responsiveness **IR03** Average resolution time per incident severity |
| SELT | **D:** reflects the automatic scale-up/scale-down of resources (compute, network, storage) over a given period of time and is linked to SA. **M:** burst tests or real-time monitoring of the affected resources, CPU cores and speed, memory size or network bandwidth | **SELT01** Number of simultaneous services **SELT02** Maximum resource capacity **SELT03** Service throughput **SELT04** Automatic load balancing **SELT05** Ratio of failed resource provisioning requests to the total number of resource provisioning requests over a commitment period. |
| DLCM | **D:** defines the provider's data handling practices, for example backup and restore strategy, replication and data export or data loss prevention. The SDDCs DLCM relies on backup or snapshots integrity checks or restoration speed durability. The deletion of data and reporting of data-loss-prevention systems can be included. **M:** backup time stamps and backup/restore durations | **DLCM01** Data mirroring **DLCM02** Data backup/snapshot **DLCM03** Data backup frequency **DLCM04** Retention time **DLCM05** Maximum data restoration time **DLCM06** Percentage of successful data restorations **DLCM07** Data deletion type |
| TCVM | **D:** measures the ability of a service to comply with a technical security policy and the handling of vulnerabilities. Regular scans of all affected systems can be used. Additionally, mailing lists and/or public vulnerability reporting programs should be included.<br>**M:** Monitor the time between vulnerability discovery and patch/fix appliance in relation to the class | **TCVM01** Percentage of timely vulnerability corrections<br>**TCVM02** Percentage of timely vulnerability reports<br>**TCVM03** Reports of vulnerability corrections |
| CM | **D:** monitors critical changes in security relevant properties of the system, for example the loss of certifications or major system upgrades. It relies on time-related change management processes and triggers or the time to implement security-related change requests.<br>**M:** using the main-stream supports duration, count applicable security-related certifications | **CM01** Critical software change reporting notifications<br>**CM02** Percentage of timely software change notifications<br>**CM03** Software security related certificates |
| DI | **D:** covers discrete access to a shared pool of resources by legitimate users for legitimate purposes and must be present at all time. Additionally, cryptographic quantitative metrics can be included, for example key lengths or hash algorithms.<br>**M:** penetration tests | **DI01** User authentication level<br>**DI02** Authentication<br>**DI03** User access storage protection<br>**DI04** Cryptographic brute force resistance<br>**DI05** Key access control policy |
| LMF | **D:** covers the access to information about historical events related to the usage of the service allocated to the customer<br>**M:** regular randomized tests of log availability | **LMF01** Log access availability<br>**LMF02** Log retention period |

## V. Assessment of IAAS Platforms

This paper focuses on the comparison of two well-known IaaS cloud platforms, industrial and open source, concerning security related SLOs. Therefore, the VMware vCloud Suite will be assessed and compared with the OpenStack software, including its components and services, based on security related SLOs extracted from the ENISA guideline.

### A. VMware

VMware is one of the world's leading suppliers for x86 server virtualization infrastructure [17]. This paper uses VMware's cloud management platform vCloud Suite. The VMware vCloud Suite (vCloud) includes VMware's industry hypervisor vSphere and vRealize, the hybrid cloud management platform for providing a SDDC [18]. VMware's vCloud Suite comprises virtualization technologies for computing, network, storage and management tools for providing SDDC capabilities to the customer.

Within this section the assessment of the applicability and implementation of extracted security related SLOs for each P.G., as listed in Table I, is described. The assessment addresses the following three VMware products: (i) the free hypervisor vSphere Hypervisor (ESXi), (ii) the industry hypervisor vSphere Suite (vSphere), and (iii) the whole VMware vCloud Suite (vCloud).

**SA:** ESXi offers an interface for monitoring the Service Availability. The measurement of response times (SA04) or redundancy (SA06) is not possible. vSphere and vCloud provide measurement and reporting tools for these parameters.

**IR:** ESXi is a free hypervisor, but does not offer SLO related support for Incident Response. In short, for ESXi this parameter is not applicable. vSphere and vCloud have different support contracts addressing this parameter.

**SELT:** ESXi allows a manually scale-up and scale-down of resources for virtual machines. Service elasticity or (automatic) load tolerance is not available. vSphere and vCloud support service elasticity and load tolerance for both, the virtual machine and the underlying hardware. High availably is also provided by vSphere and vCloud. Tenant specific resource management, capacity and performance management (SELT04, SELT05) are provided by vCloud only. The measurement of resources is available in all three products (SELT03).

**DLCM:** ESXi does not provide access to the storage API. This API is used by backup applications for snapshot management. Therefore, ESXi does not offer DLCM. Snapshots are available in all three products (DLCM02), both vSphere and vCloud provide comprehensive measurement and reporting tools regarding this parameter.

**TCVM:** In general hotfixes and patches are available for all three products. The Security Response Policy offered by VMware states the correction of security issues in a timely fashion (TCVM03). Integrated patch management is provided by vSphere and vCloud, compliance measurement is provided by vCloud only.

**CM:** ESXi does not offer tools regarding software change management. Major upgrades are installed manually. vSphere offers basic tools regarding change management, like software deployment tools for upgrades. vCloud provide comprehensive measurement and reporting tools supporting software change processes (CM01).

**DI:** ESXi does not provide DI in a multi-tenant environment. vSphere uses pools for data isolation within one security boundary. vCloud offers data isolation in a multi-tenant environment. All three products use secure connections for resource access. Different types of user authentication and user rights management (DI01, DI02) are offered by vSphere and vCloud only.

**LMF:** ESXi offers basic log access (LMF01). vSphere and vCloud provide Log Management, supporting different formats, achievement and access on user basis.

To avoid redundancy, the results of the analysis of the applicability and implementation of ENISA parameters on VMware are only presented in the comparison in Table III.

## B. OpenStack

OpenStack is an open-source platform for cloud computing providing IaaS. The platform consists of separate modules, where each offers a specific service, and acts as a software-defined-datacenter. The interaction of all OpenStack services enables a high integration to provide IaaS within SDDC. Within this section the assessment of the applicability and implementation of extracted security related SLOs for each P.G., as listed in Table I, is described.

**SA:** A single-controller high availability mode, which is managed by the services that manage highly available environments, is supported by OpenStack (SA01). Redundant controllers for failovers are not configured (SA06).

**IR:** Currently not applicable on OpenStack.

**SELT**: Load balancing as a service (SELT04) is offered by Neutron. Furthermore, Octavia provides additional capabilities for load balancing (including the usage of compute drivers to build instances which operate as load balancer).

**DLCM:** Cinder services can be used for management of volumes and snapshots (DLCM02) for use with the Block Storage API. The self-service API supports end users to request and consume resources.

**TCVM:** OpenStack offers a vulnerability management process, but the manual installation of hotfixes and patches by root is necessary (TCVM03).

**CM:** OpenStack architecture and processes are documented on the community webpage, but there is no automatism to inform the cloud service customer (CM01). Furthermore, there are no supporting change management services within OpenStack.

**DI:** OpenStack provides a separation of tenants. With the implementation of OpenStack's Identity API, Keystone provides API client authentication, service discovery and distributed multi-tenant authorization (DI02).

**LMF:** OpenStack provides standard logging levels depending on increasing severity. The possible levels are "debug", "info", "audit", "warning", "error", "critical" and "trace". Horizon offers the dashboard for the end user, with limited access to the content of the log-files (LMF01). A Log Management Data Flow by Monasca is provided.

To avoid redundancy, the results of the analysis of the applicability and implementation of ENISA parameters on OpenStack are only presented in the comparison Table III.

## C. Results and Discussion

To compare the applicability of SLOs, as described in the previous sections, the following grades have been defined:

TABLE II. GRADES OF APPLICABILITY

| Grade | Applicability of SLO |
|---|---|
| F | The SLO is fully applicable and natively implemented. |
| P | The SLO is implemented, but cannot be used without additional administrative tasks, e.g.: a feature is natively implemented, but needs to be installed manually (partially applicable and implemented). |
| N | The SLO is not applicable or not implemented. |

After assessing both VMware and OpenStack, these grades have been used to point out whether the applicability and implementation of ENISA security P.G.s were fully applicable (F), partially applicable (P) or not applicable (N). Table III summarizes the results of the comparison:

TABLE III. COMPARISON OF APPLICABILITY AND IMPLEMENTATION OF ENISA PARAMETER GROUPS BETWEEN OPENSTACK AND VMWARE PRODUCTS

| ENISA parameter group | OpenStack | VMware | | |
|---|---|---|---|---|
| | | vCloud | vSphere | ESXi |
| SA | P | F | F | P |
| IR | N | F | F | N |
| SELT | F | F | P | N |
| DLCM | P | P | P | N |
| TCVM | P | F | P | P |
| CM | P | F | P | N |
| DI | F | F | P | N |
| LMF | P | F | F | P |

As presented in Table III, the results of the analysis show a high level of applicability and implementation of ENISA SLOs in VMware's vCloud Suite. Nearly all ENISA parameters are fully covered by vCloud. The major gap is the lack of Data Loss Prevention. In contrast, OpenStack and its services fulfill most of the ENISA criteria partially.

The results show two main outcomes:
- OpenStack does not address the P.G. of IR, whilst in vCloud and vSphere it is natively implemented and applicable.
- In both, VMware and OpenStack the DLCM is only partially implemented.

Hence the contribution is twofold, first we identify a set of topics that still require research and development (e.g.: IR, DLCM) and secondly, as a practical output, we provide a comparison of popular industrial and open-source platforms focusing on private cloud environments, which are important for Industry 4.0 use cases.

## VI. CONCLUSION AND FUTURE WORK

In this work, due to the broad applicability of ENISA, the focus lies on the evaluation of ENISA guidelines concerning security assessment of IaaS solutions. The extracted set of security parameters is used for the assessment of the industrial VMware platform and the open source OpenStack platform.

The results of the assessment have shown a high applicability of ENISA criteria on the VMware vCloud, where only one P.G. is met partially. In contrast to VMware vCloud, OpenStack fulfills most of the parameter groups only partially, but only one is not fulfilled at all. Compared to the existing work, the evaluation shows another way for assessing IaaS platforms. The results show the different levels of applicability of ENISA criteria on an industrial platform, where we expected a lot of features, and on an open source platform, which provides many modules from the open source community.

The results are a first step towards the technical definition of a security baseline for IaaS cloud back-ends, which can be used for Industry 4.0 environments. Furthermore, we have identified a lack of support of IR and of DLCM in IaaS platforms, suggesting a need for research in these topics. To confirm this, future work will investigate current cloud research projects, to find out how well the research community supports this Industry 4.0 related issues. Additionally, we will include a wider set of guidelines to create a more comprehensive catalog.

ACKNOWLEDGMENT

This work has been performed as part of the project Power Semiconductor and Electronics Manufacturing 4.0 (SemI40), under grant agreement No 692466. The project is co-funded by grants from Austria, Germany, Italy, France, Portugal and - Electronic Component Systems for European Leadership Joint Undertaking (ECSEL JU).